\documentclass[a4paper,fleqn,useAMS,usenatbib]{mnras}
\usepackage[T1]{fontenc}
\usepackage{ae,aecompl}
\usepackage{graphicx}
\usepackage{bm}
\usepackage{graphics}
\usepackage{color}
\usepackage{epstopdf}
\usepackage{amsmath}

\usepackage{times}
\usepackage{natbib}
\usepackage{setspace}
\usepackage{hyperref}
\usepackage{amsbsy}
\usepackage{gensymb}

\newif\ifAMStwofonts
\AMStwofontstrue

\def\apj{ApJ}                 
\def\apjl{ApJ}                
\def\apjs{ApJS}               
\def\ao{Appl.~Opt.}           
\def\aap{A\&A}                
\def\mnras{MNRAS}             
\def\prd{Phys.~Rev.~D}        
\def\nat{Nature}              
\def\jcap{JCAP}

\makeatletter
\newcommand\lsim{\mathrel{\rlap{\lower4pt\hbox{\hskip1pt$\sim$}}
\raise1pt\hbox{$<$}}}
\newcommand\gsim{\mathrel{\rlap{\lower4pt\hbox{\hskip1pt$\sim$}}
\raise1pt\hbox{$>$}}}

\newcommand{\bc}{\mathrm{bc}}
\newcommand{\msun}{M_\odot}

\usepackage{color}

\makeatother

\title[Spin parameters of the first objects]{The Supersonic Project: rotational effects of supersonic motions on the first structures in the Universe}

\author[Chiou et al.]{Yeou S. Chiou$^{1,2}$\thanks{yschiou@physics.ucla.edu },  Smadar Naoz$^{1,2}$,  Federico Marinacci$^{3,4}$, Mark Vogelsberger$^{3}$
\\
$^{1}$Department of Physics and Astronomy, University of California, Los Angeles, CA 90095\\
$^{2}$Mani L. Bhaumik Institute for Theoretical Physics, Department of Physics and Astronomy, UCLA, Los Angeles, CA 90095, USA\\
$^{3}$Department of Physics and Kavli Institute for Astrophysics and Space Research, Massachusetts Institute of Technology, Cambridge,\\ MA 02139, USA\\
$^{4}$Harvard-Smithsonian Center for Astrophysics, 60 Garden Street, Cambridge, MA 02138, USA
}

\date{Accepted 2018 September 06. Received 2018 September 06; in original form 2018 July 16.}
\pubyear{2018}

\begin{document}
\label{firstpage}
\pagerange{\pageref{firstpage}--\pageref{lastpage}}
\maketitle

\begin{abstract}
We introduce the ``Supersonic Project,'' aimed at investigating the effects of the supersonic relative velocity between dark matter (DM) and  baryons at high redshift using a combination of analytical calculations and cosmological simulations. In this paper, we study the effect of this stream velocity on the angular momentum of the first structures in the early Universe using simulations. We focus on DM haloes and their gas component as well as the recently predicted supersonically-induced gas objects (SIGOs) that arise as a result of the stream velocity phase shift. We find that the spin parameter of the gas component in these first haloes is increased with the stream velocity. Moreover, we find that when the stream velocity is taken into account, the angular momentum vectors of the DM component and the gas component are typically misaligned and this misalignment angle has a nearly isotropic distribution. The spin parameter value of the gas component is higher than in the no stream velocity case, which even in the absence of cooling, may result in more prolate objects. 
We also generalize the spin parameter to the SIGOs and find that they typically have a larger spin parameter with respect to their dark matter counterparts and that there is no correlation of the spin parameter and the prolateness of such structures. We speculate that SIGOs may be observed as very low luminosity objects in the early Universe and may serve as potential progenitors of Little Blue Dot-like systems. 
\end{abstract}

\begin{keywords}
cosmology: theory -- methods: numerical -- galaxies: high redshift
\end{keywords}

\section{Introduction}
The Lambda-Cold Dark Matter model ($\Lambda$CDM) has had great success in characterizing structure formation and the evolution of the Universe. Using this model, numerical simulations have been able to accurately describe structure on the largest of scales \citep[$\sim (100$~Mpc$)^3$, e.g.,][]{springel05,vogelsberger14a,vogelsberger14b,schaye15}.
The currently accepted paradigm suggests that present-day galaxies evolved from primordial inhomogeneities in the density distribution at very early times \citep[e.g.,][]{rees77,silk77,white78,blumenthal84}.
Dark Matter (DM) over-densities had been evolving logarithmically with the scale factor during the radiation-dominated era and linearly during matter-dominated era. However, baryons decoupled from the radiation only at the time of recombination, and thus were free to start growing \citep[e.g.,][]{Padmanabhan}. This resulted in a significant difference between the over-density amplitudes of DM and baryons at the time of recombination \citep{NB}, and in a relative velocity between the two components as well. \citet{TH} showed that the relative velocity between these over-densities was supersonic at the time of recombination as a consequence of the sudden drop in baryonic temperature. In particular, it had an rms velocity of $\sim 30$~km$~\text{s}^{-1}$  and was coherent on scales of less than a few comoving Mpc. This effect was coined the  ``stream velocity'' effect. In terms of cosmological perturbation theory, the relative velocity coupling to the over-densities is formally a second-order effect and is often neglected in analytical work and in simulations. 

Although second-order, the supersonic nature of the stream velocity has been shown to have non-trivial effects on the evolution of the over-densities within those patches such as lowering the number density of haloes \citep[e.g.,][]{TH,miao11,stacy11,tseliakhovich11,fialkov2012,naoz12,oleary12,bovy13,tanaka13,tanaka14,asaba16}. 
\citet{naozyoshida13}, showed that  the stream velocity also increases the minimum halo mass that is capable of retaining most of its baryons throughout its formation epoch, yielding first structures that are deprived of gas.  
Overall, the gas fraction in the first DM haloes decreases  \citep[e.g.,][]{dalal10,greif11,miao11,tseliakhovich11,fialkov2012,naoz12,naozyoshida13,oleary12,richardson13,asaba16}.
As a result, star formation in haloes is delayed because haloes are not capable of accreting enough gas at early times \citep[e.g., $z\lsim 10$][]{liu11,miao11,greif11,fialkov2012,oleary12,bovy13,kyungjin,richardson13}.  It was also shown that the stream velocity, in combination with spatial fluctuations in the speed of sound \citep{NB}, can cause the formation of coherent large-scales magnetic fields \citep{naoznarayan13}. Moreover, it was suggested that the delayed star formation can result in black hole formation via direct collapse and largely affect primordial black hole  evolution \citep[e.g.,][]{tanaka13,tanaka14,latif14,schauer17,hirano17}. Finally, the stream velocity has important implications for 21-cm cosmology since the stream might enhance the hydrogen signal from the different patches in the sky \citep[e.g.,][]{visbal12} and might also be related to the trough in 21-cm via DM-baryon scattering \citep{barkana2018,barkana+18}.

\citet{naoznarayan14}  demonstrated, analytically, that the supersonic nature of the relative velocity results in a physical off-set between the DM and baryonic over-densities. This leads to the formation of Supersonically-Induced Gas Objects (SIGOs) outside the virial radius of dark matter haloes. They conjectured that these structures could be the progenitors of globular clusters, which tend to be devoid of DM. Additionally, the parent gas-deprived dark matter haloes could potentially explain some of the dark satellite galaxies.
\citet{popa}, using simulations performed with the moving-mesh code {\sc arepo} \citep{springel10}, revealed that these SIGOs do indeed exists outside the virial radius of DM haloes. SIGOs also tend to have an ellipsoidal structure, in contrast to the roughly spherical nature of the DM haloes. Recently, \citet{hirano17b} also confirmed the existence of these structures in a cosmological simulations using the SPH code {\sc gadget-2} \citep{springel05}. Utilizing a zoom-in method, they followed the fragmentation of gas clouds into SIGOs and suggested that these clumps are the progenitors of the first star clusters.  

SIGOs are therefore of great interest; they might be the first star clusters and/or represent the present day globular clusters. One potentially promising observable of these SIGOs in particular, and the stream velocity as a whole, is the angular momentum of these objects. Both SIGOs and DM haloes are affected by the stream velocity. Thus, one expects a notable signature on their angular momentum and specifically on the so-called spin parameter, which is a normalized angular momentum. The angular momenta of the DM and the gas component of haloes are directly related to their formation history \citep[e.g.,][]{fall80,mo,peebles}. Within the context of the hierarchical structure formation picture, baryons lose their angular momentum to the DM as they spiral into the centre. In the presence of the stream velocity, there is a preferred directionality that is expected to increase the spin parameter.
 
Furthermore, the angular momentum distribution of an object is correlated with the formation of a gaseous disk. Earlier studies of disk formation typically assume that early tidal torques endow the same angular momentum to DM and baryons \citep[e.g.,][]{barnes87,fall80, mo}. However, such an assumption breaks down when considering the non-linear Universe. The inclusion of baryonic processes and feedback further complicates the problem. Given the importance of the topic, several studies have investigated the angular momentum distributions of DM haloes and baryons and have had great success in describing disk formation \citep[e.g.,][]{vandenbosch02,maller02,teklu15,RG}. Since the spin parameter gives a measure of the rotational support of an object, it was suggested that this quantity may serve as an observational signature of the parent halo at low redshift for high-mass haloes \citep[e.g.,][]{Teklu+15,Zavala+16,sommerville2017}. In particular, \citet{sommerville2017} showed that ratio of the half-light radius to the virial radius of the halo is proportional to the DM spin parameter value. However, as pointed out by \citet{Jiang+18}, there is still no clear correlation between the  DM spin parameter and the corresponding gas component spin parameter.
\citet{sommerville2017} found that the stellar radius-halo radius correlation is roughly independent of stellar mass at $z\sim 1$, but at higher redshift ($z \sim 2$) there is a mass dependence with high mass galaxies having a 50\% smaller ratio of stellar radius to halo radius than low mass galaxies. This correlation informs on the way that galaxies at high redshift interact. In particular, major mergers and violent disk instabilities may drive the loss of angular momentum from these haloes. However, observations in high redshift regimes ($z \gsim 10$) are beyond current technological reach. Detailed numerical simulations can provide the necessary insight into the complex physical processes that take place at these high redshift regimes. 
 
The goal of the ``Supersonic Project'' is to understand the stream velocity effect and how it affects structure formation in the early Universe. Future papers in this series will address the stream velocity effect with star formation and feedback prescriptions taken into account. In this first paper of the ``Supersonic Project'', we specifically investigate the consequences of the stream velocity at high redshift $(20 < z < 10)$ on the morphology and spin of DM and gas inside haloes, as well as the SIGOs, using  a suite of {\sc arepo} simulations \citep[see][]{popa}.

This paper is organized as follows. In section \ref{sec:numerics}, we briefly describe the simulation setup. We review the spin parameter and the numerical methods that were employed in section \ref{sec:struc}. We then present our results about the effects of the stream velocity on the spin parameter and misalignment angle between gas and DM for haloes in our simulations in section \ref{sec:results}. Finally, we wrap up with a summary and conclusions in section \ref{sec:conc}. 

Throughout the paper we assume a $\Lambda$CDM cosmology with $\Omega_{\Lambda} = 0.73$, $\Omega_m = 0.27$, $\Omega_B = 0.044$, $\sigma_8 = 1.7$, and $h = 0.71$. All the quantities that we analyse in this paper are expressed in physical units. 

\section{Numerical simulations}\label{sec:numerics}

We study the {\sc arepo} simulations presented in \citet{popa}. The transfer functions of the initial conditions are calculated via a modified {\sc cmbfast} code \citep{seljak96} to take into account the first-order correction of the scale dependent temperature fluctuations \citep[following][]{NB}. These corrections were shown to have a significant effect in determining the gas fraction in haloes at high redshift \citep[e.g.,][]{naoz09,naoz11,naozyoshida13}. The transfer functions also include the stream velocity evolution \citep[these are second-order corrections to the equations][]{TH}. We chose to work with the $2$~Mpc box size presented in \citep{popa}. The box has $512^3$ DM particles with associated mass resolution of $m_{\rm DM} = 1.9\times10^3~M_{\odot}$. The gas component is represented by $512^3$ Voronoi mesh cells which corresponds to a particle mass of $m_{\rm gas} = 360~M_{\odot}$. These simulations include only gravitational physics and non-radiative hydrodynamics, i.e., no radiative cooling or (sub-grid) galaxy formation physics (star formation, stellar feedback, etc.) is considered. The box was run from $z=200$ to $z=0$ twice, once with no stream velocity, and once with stream velocity $v_{bc}=2~\sigma_{v_{bc}}=11.8~\rm km~s^{-1}$, where $\sigma_{v_{bc}}$ is the rms velocity. Stream velocity values of $2~\sigma_{v_{bc}}$ were chosen because such values are able to produce gas-rich structures outside the virial radius of DM haloes with masses near that of globular clusters \citep{naoznarayan14}. Both DM and gas structures in \citet{popa} were identified using the Friends-of-Friends (FOF) algorithm with a linking length of $0.2$ times the mean particle separation to identify haloes and their corresponding virial radius. Following \citet{popa} we defined two different structures with the FOF algorithm:
\begin{enumerate}
\item \emph{DM-Primary/Gas-Secondary} objects are those for which FOF was run on the DM first. Gas cells were then linked in a secondary stage \citep{dolag09} in order to investigate gaseous substructure in the DM haloes. A spherical overdensity (SO) calculation is performed to obtain the virial radius from these FOF groups. 
\item \emph{Gas-Primary} objects are found by running the FOF algorithm only on the baryonic component\footnote{One important question is if these objects are bound. These objects may be falling into a bigger potential well. Furthermore, numerical effects (see below) tend to disperse these objects.}. The Gas-Primary structures tend to be more filamentary in nature, thus, we adopt the ellipsoidal fitting algorithm from \citet{popa} to each FOF group. Since some of these structures tended to be very sparse, we implement a ``tightly fitting ellipsoid'' to each object in the following manner: keeping the axis ratios constant, the ellipsoid is progressively shrunk in decrements of $0.5\%$ until the the ratio of the lengths of the axis of the shrunken ellipsoid to the original is greater than the ratio of the number of gas cells in the shrunken ellipsoid to the original or until $20\%$ of the particles have been removed. Note that many of these Gas-Primary structures reside inside DM halo, as they are simply the gas components of the DM-Primary/Gas-Secondary structures. 
\end{enumerate}
Our definitions of structures are summarized in Table \ref{definitions}. In the analysis that follows, we focus on the morphology and angular momentum of these high-redshift $(20 < z < 10)$ structures.

\begin{table*}
\centering
\begin{tabular}{|p{5cm}||p{5cm}||p{5cm}|}
\hline
DM-Primary/Gas-Secondary & Gas-Primary & Supersonically Induced Gas Objects (SIGOs) \\ \hline\hline
Objects for which FOF was run on the DM first, then gas was linked in a secondary stage & Objects for which FOF was run on the gas only & Gas-Primary objects that are outside the virial radius of the closest DM halo and have gas fractions greater than 40\% \\ \hline
\label{definitions}
\end{tabular}
\caption{Table of structure definitions used in the paper}
\end{table*}

\section{Structure and Morphology}\label{sec:struc}

\begin{figure*}
\includegraphics[width=.75\textwidth]{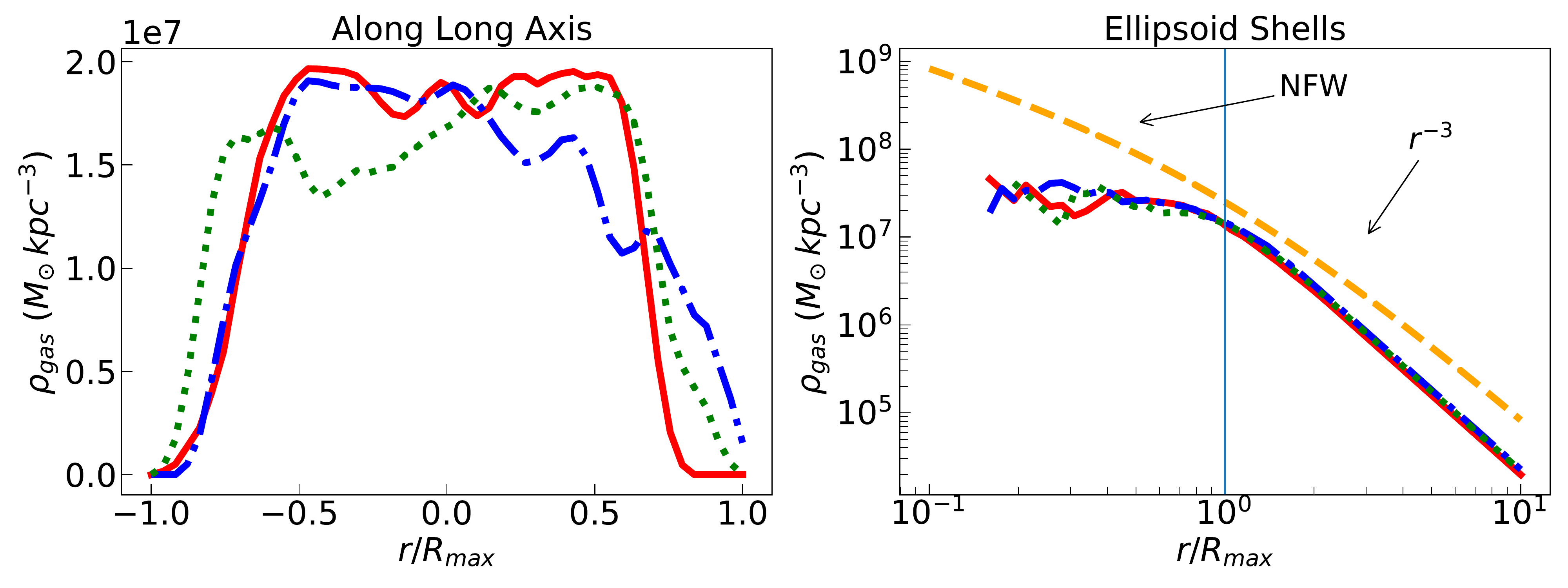}
    \includegraphics[width=.33\textwidth]{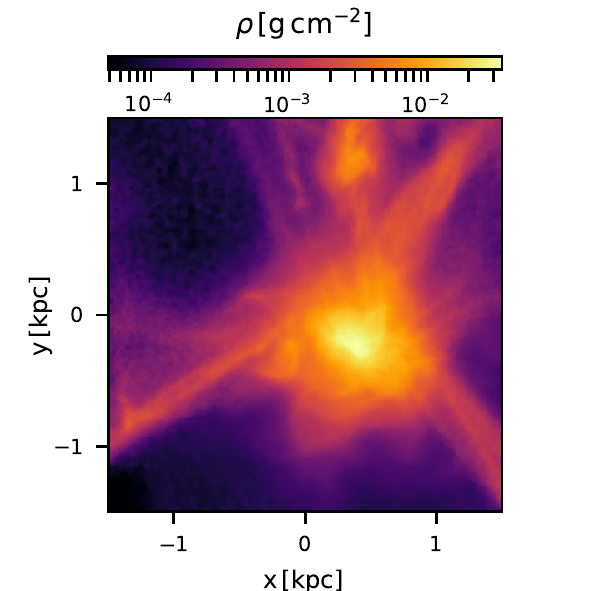}\hfill
    \includegraphics[width=.33\textwidth]{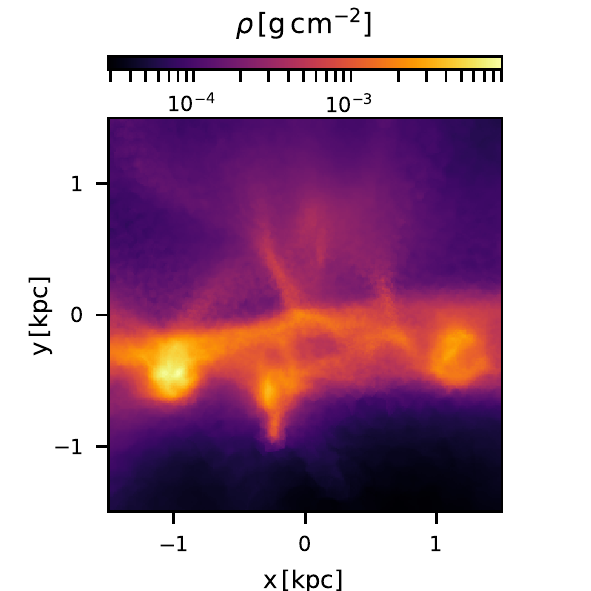}\hfill
    \includegraphics[width=.33\textwidth]{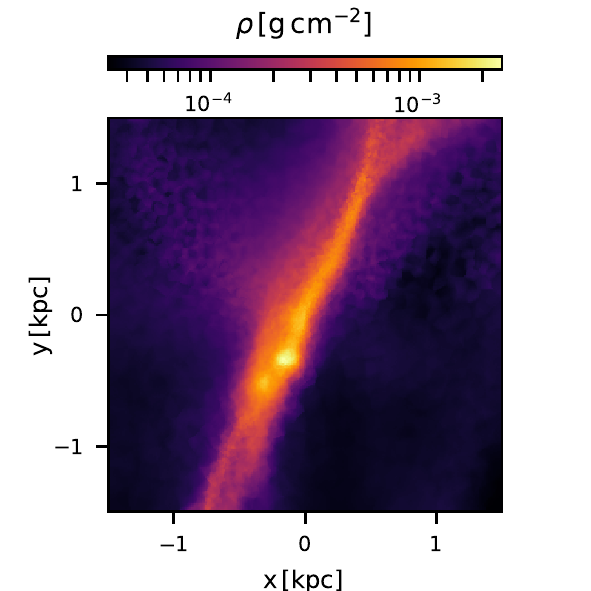}
    \caption{{\bf Top row:} {\it Three examples of the density profiles of Gas-Rich Objects} The left panel depicts the density along the long axis of ellipsoid, while the right panel shows the density in increasing ellipsoidal shells. For illustration purposes, in the right panel, we present a typical NFW profile, with $\rho_0=10^8$~M$_\odot$~kpc$^{-3}$, and replacing $r/r_s\to r/R_{\rm max}$ (dashed orange line). The left panel densities have been smoothed with a moving average filter. {\bf Bottom row:} Projected density profiles (in $\rm{g~cm^{-2}})$ for the three example ellipsoids from the top row at $z=20$. In particular, the left panel corresponds to the red (solid) line in the top row, while the middle panel corresponds to the green (dotted) line in the top row, and finally the right panel corresponds to the blue (dot-dashed) line in the top row.}
    \label{gasproj}
\end{figure*}

The spherically averaged density profiles of DM haloes are typically well described by an NFW \citep{navarro96} profile. One can associate a radius, $R_{200}$, to these haloes, which corresponds to the radius of a sphere with overdensity 200 times the critical density for closure of the Universe\footnote{It has been suggested that $R_{200}$ does not accurately describe a physical boundary for a halo. In particular, processes such as further mass accretion may cause the actual radius of the halo to be larger than $R_{200}$. This was the motivation for the ``splash back'' radius \citep{More2015,diemer17}. Although there are possible disadvantages of $R_{200}$, here we us it as a simple normalization which allows us to compare to previous studies.}. Note that DM haloes show clear triaxiality \citep[e.g.,][]{Sheth+01,Lithwick+11,Vogelsberger+11,Schneider+12}, and are assumed to be spherically symmetric here for simplicity. 
As depicted in Figure 8 of \citet{popa}, the NFW profile is preserved even with the introduction of the stream velocity.  In contrast to the DM profile, \citet{popa} showed that Gas-Primary components tend to be more ellipsoidal in nature.  Typical density profiles for Gas-Primary objects are displayed in Figure \ref{gasproj}, top row. Estimating the density in ellipsoid shells demonstrates that the profile deviates from an NFW profile for these objects. We also consider density profiles as a function of the longest axis of the ellipsoid which describes the concentration of mass along the axis. We have smoothed the density along the longest axis using a moving average filter to minimize particle discretization effects. We find three typical cases. The first, shown in solid red in Figure \ref{gasproj}, demonstrates a relatively flat distribution in the middle of the ellipsoid corresponding to the left panel in the bottom row of Figure \ref{gasproj}. The second, shown in dotted green, exhibits multiple peaks, which might suggest a merger or a breakup, corresponding to the middle panel in the bottom row. Lastly the third, in dot-dashed blue, shows an increasing density towards one direction, corresponding to the right panel in the bottom row. This could be caused by gravitational attraction to the closest dark matter halo. At large distances (i.e., $r>R_{\rm max}$ for SIGOs and $r>r_{s}$ for DM haloes, where $r_s$ is the scale radius of NFW) the distribution drops sharply. However, the inner slope is different, and tends to be flatter for the Gas-Primary objects. The gas density projections (Figure \ref{gasproj}, bottom row) show asymmetric gas distributions inside the these objects.

To properly analyze SIGOs, we will define them to be Gas-Primary structures whose centre of mass lies outside the virial radius of the closest dark matter halo and that have gas fraction $> 40\%$\footnote{The choice of $40\%$ is somewhat arbitrary, we note that it is above the cosmic baryon fraction and close to $50\%$, i.e., the stellar fraction of globular clusters \citep{heggie96,bradford11,conroy11,ibata13}}.

\begin{figure} 
\includegraphics[width=.45\textwidth]{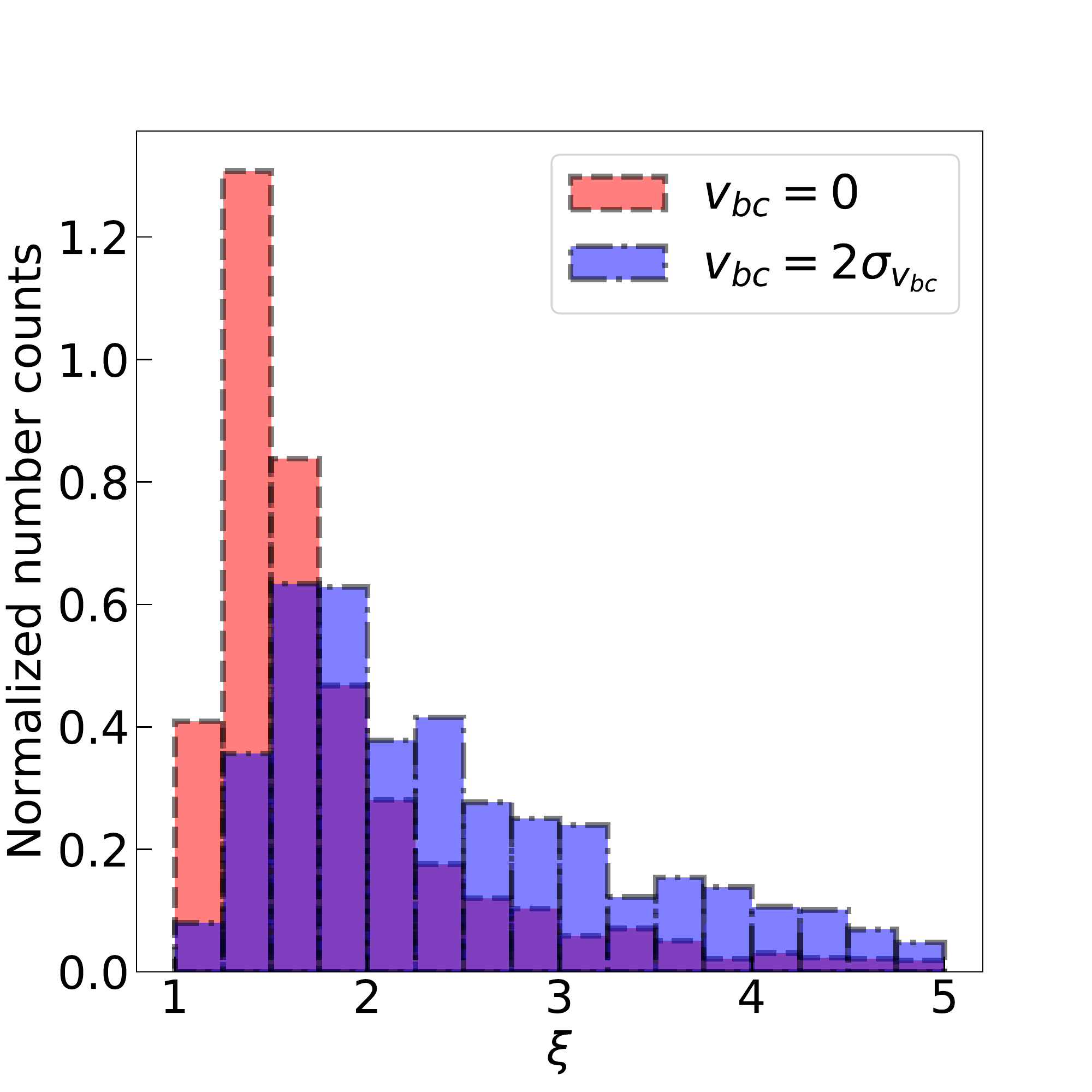}
\caption{A histogram of the prolateness factor, $\xi$ (see Eq.~\ref{eq:pro}) for Gas-Primary objects at $z=20$. The zero stream velocity case is shown in dashed red, whereas the $v_\bc=2\sigma_{v_{\rm bc}}$ in dot-dashed blue. The stream velocity shifts the prolateness of Gas-Primary objects towards higher values.}
\label{rmaxrminhist}
\end{figure}

To characterize the ellipsoidal morphology of Gas-Primary objects, we define a prolateness factor as 
\begin{equation}\label{eq:pro}
    \xi =\frac{R_{\rm max}}{R_{\rm min}} \ ,
\end{equation} 
where $2R_{\rm max}$ ($2R_{\rm min}$) refers to the length of the maximum (minimum) axis of the ellipsoid. In Figure \ref{rmaxrminhist} we show the distribution of $R_{\rm max}/R_{\rm min}$ for our gas ellipsoids with and without the stream velocity. As can be seen in this Figure the stream velocity tends to produce more prolate systems. 

\begin{figure}
\includegraphics[width=.5\textwidth]{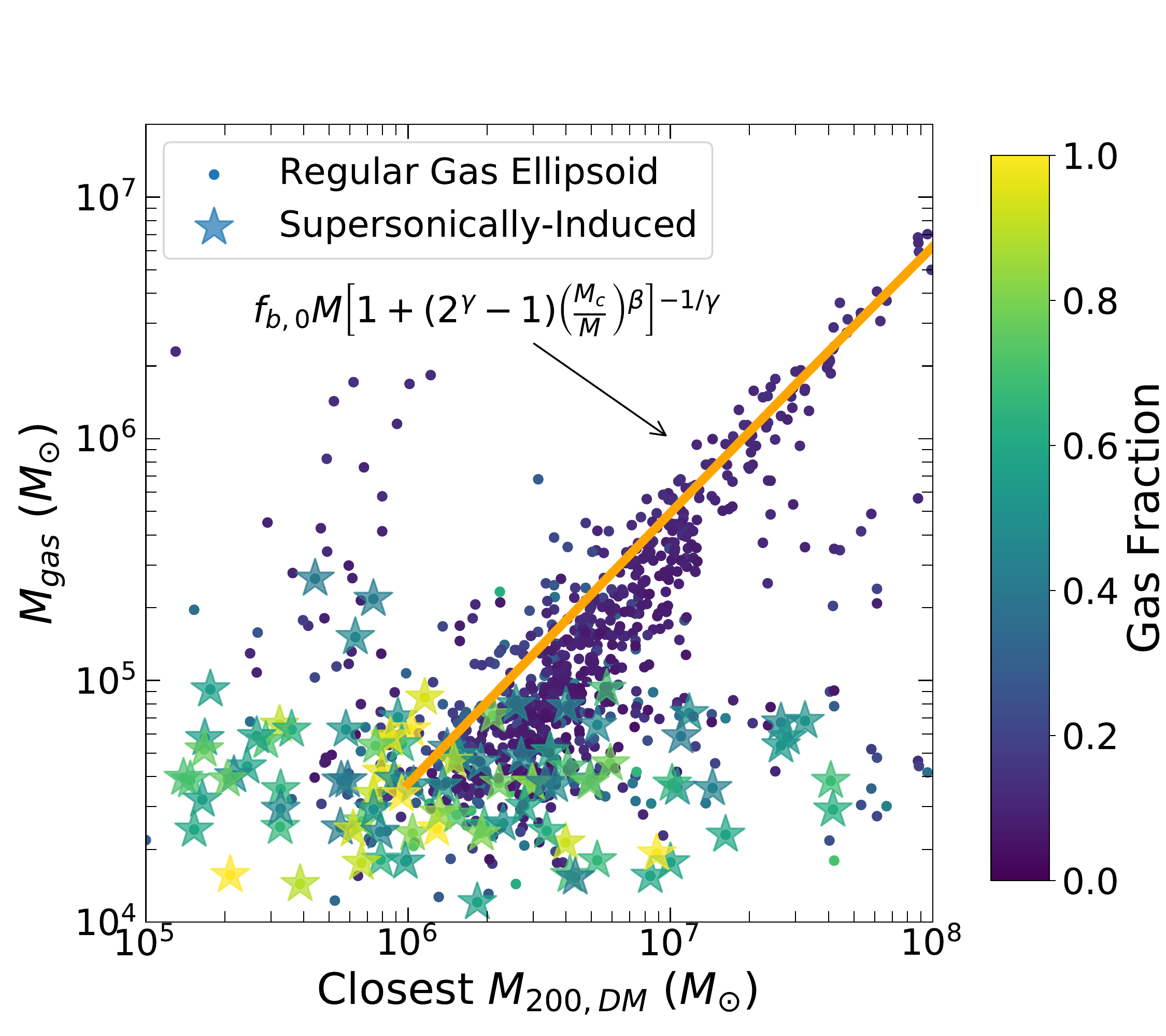}
\caption{Comparison of the gas mass inside ellipsoids and the virial mass of the closest dark matter halo at redshift $20$ for the nonzero stream velocity run. Stars refer to SIGOs. The orange line refers to the fit to gas mass derived from an initial DM halo (see Equation \ref{eq:Mc}). This characteristic mass fit describes the fact that smaller DM haloes will have less  gas. The fit parameters are $\gamma=8$, $\beta=1$, $M_c=6\times 10^5~M_{\odot}$, and $f_{b,0} = 0.07$. We fit only objects that are not SIGOs.
}
\label{closestm200} 
\end{figure}

\citet{naoznarayan14} used analytical calculations to show that SIGOs originated from a corresponding DM over-density mode that is shifted by a physical distance which can be larger than the virial radius of the DM halo. Thus the gas-deprived parent DM halo (i.e., the fiducial halo corresponding to the DM over-density mode) simply has a mass of $M_{\rm  DM} \sim 6 M_{\rm gas}$ since DM is on average $5$ times as massive as gas. One might naively expect that the closest DM halo of a SIGO will be the parent halo. However, due to dynamical interactions, we find that the closest DM halo is not necessarily the host halo, as can be seen in Figure \ref{closestm200}. In other words, SIGOs move around and tend to cluster around more massive DM haloes. In Figure \ref{closestm200}, SIGOs are represented by star symbols and show no correlation with $M_{\rm gas}$. 

The suppression of the gas component in DM haloes from the stream velocity effect can be evaluated analytically from the power spectrum. The typical mass at which the gas fraction is substantially suppressed, when estimated from such calculations (for both $v_\bc=0$ and $v_\bc\neq0$), agrees with the fits from simulations  \citep[the so called characteristic or filtering mass, e.g.,][]{gnedin00,Naoz+07,naoz09,naoz11,naozyoshida13,popa}.
Thus, we can find a simple correlation between the gas mass in a DM halo and the DM mass of that halo, using the characteristic mass $M_c$:
\begin{equation}\label{eq:Mc}
M_{\rm gas} = f_{b,0}M\left[1+(2^{\gamma}-1)\left(\frac{M_c}{M}\right)^{\beta}\right]^{-1/\gamma} \ ,
\end{equation}
 where we adopt the following fit parameters: $\gamma=8$, $\beta=1$, $M_c=6\times 10^5$~$M_{\odot}$ \citep[e.g.,][]{popa}. Note that $f_{b,0}$ is defined as the gas fraction of the top $5\%$ massive haloes realized. For the nonzero stream velocity run, we have $f_{b,0} = 0.07$.  This fit is depicted in Figure \ref{closestm200} (orange line) and shows an agreement with the apparent correlation of the gas components inside their DM haloes. The SIGOs exhibit scatter and do not lie on this line and therefore do not display the classical behaviour of gas clumps inside DM haloes. 
 
Note that the relatively small number of SIGOs, with 18 (98) objects at redshift 10 (20) is probably numerical in nature. In {\sc arepo}, gas cells interact as gravitationally softened particles. Thus, the gas component in a SIGO has a velocity distribution, with a tail of particles' velocities above the escape velocity of the SIGO. These particles then become unbound, and the SIGO slowly evaporates. To estimate the timescale of this evaporation, we adopt a crude collisionless approach \citep{naoznarayan14,gnedin14,gieles11}. This collisionless process results in the unbinding (evaporation) of star clusters \citep[e.g.,][]{gieles11} as well as the low mass SIGOs in our simulation. Specifically, \citet{gnedin14}  showed that the evaporation timescale of a $2\times 10^5$~M$_\odot$ star cluster is approximately $17$~Gyr, if all the stars are $1$~M$_\odot$. From the point of view of gravitationally softened interacting particles, a $2\times 10^5$~M$_\odot$ SIGO mass object has about $555$ particles (for our mass resolution of $M_{\rm gas}=360$~M$_\odot$). The evaporation timescale for this object is $0.045$~Gyr, which is much shorter than the time passed between $z=20$ to $z=10$. Therefore our insufficient resolution results in the numerical loss of many SIGOs (which implies that the SIGOs are marginally bound). In follow up investigations we plan to use a zoom-in methodology to increase the resolution.

\section{The Angular momentum of Gas-rich and Gas-poor structures} \label{sec:results} 

\subsection{The spin parameter}

\begin{figure}
\includegraphics[width=.5\textwidth]{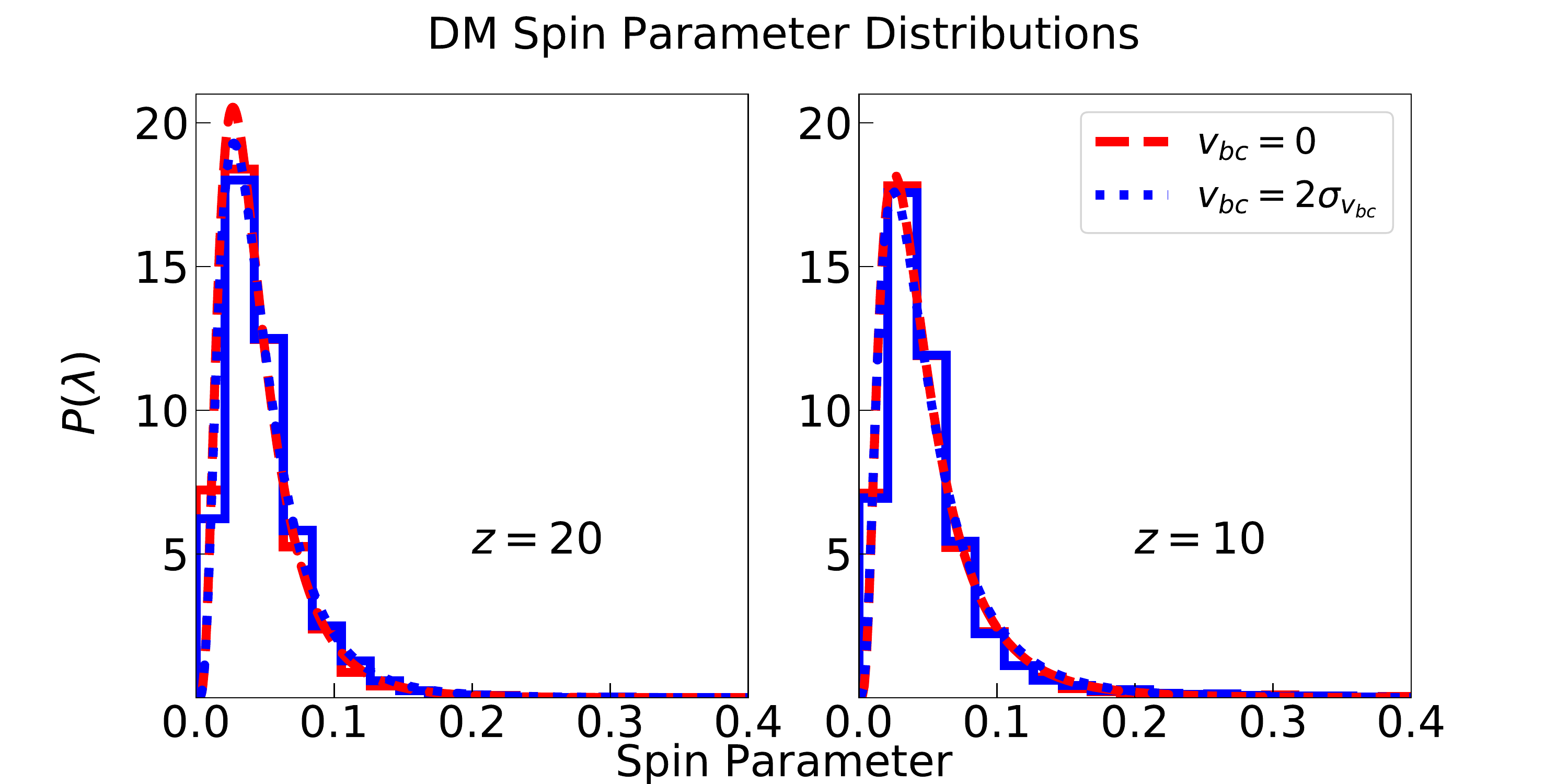}
\includegraphics[width=.5\textwidth]{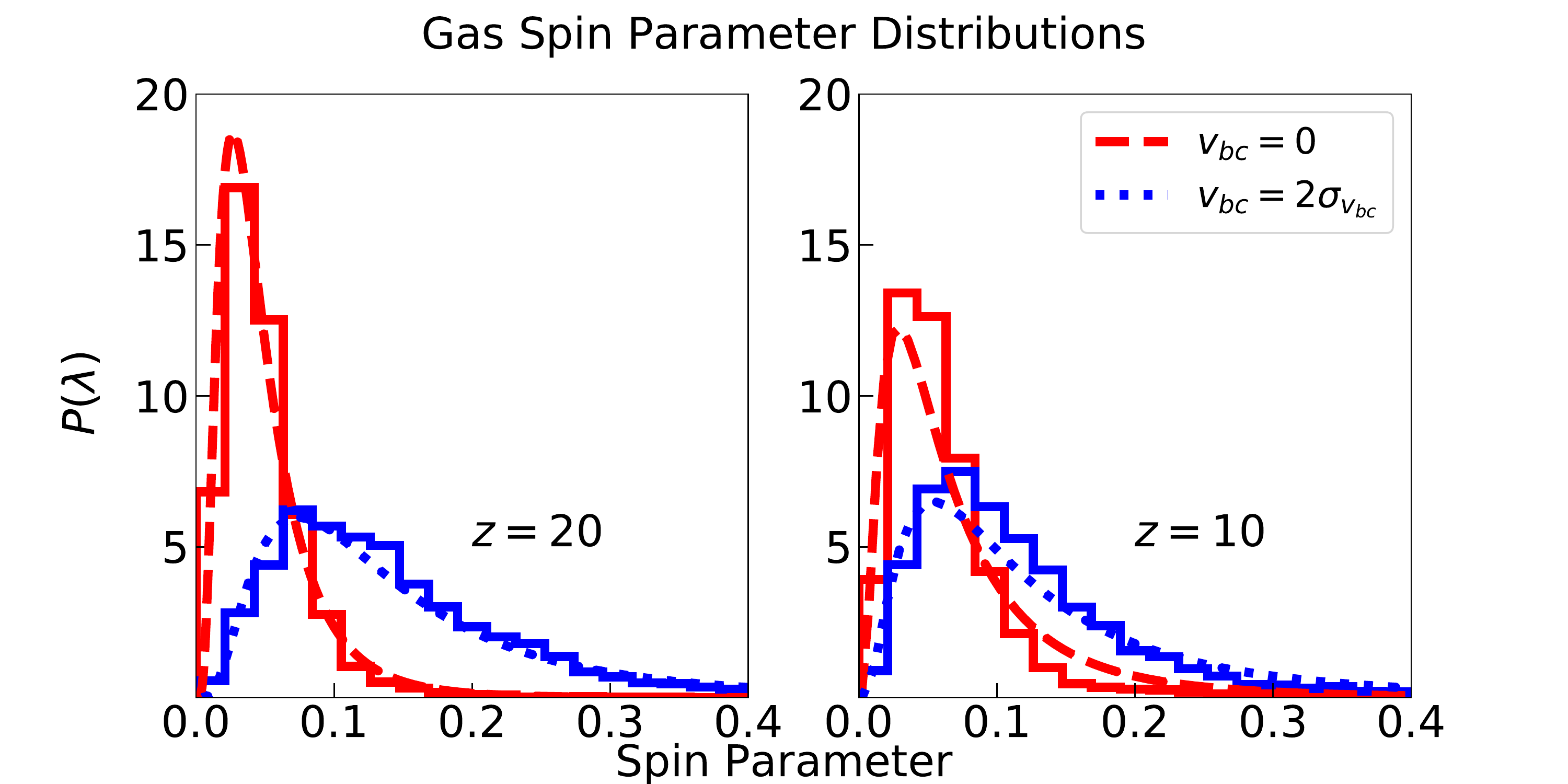}
\caption{Spin parameters distributions of DM (top panels) and gas in the DM haloes (bottom panels). We consider the zero stream velocity case (dashed red lines) and the $v_\bc=2\sigma_{\rm vbc}$ (dotted blue lines), for two redshift examples ($z=10$ and $z=20$). We also provide a lognormal fit to the distribution. The fit parameters are available in Table \ref{table:Spin}. }
\label{lognormaldist}
\end{figure}

The angular momentum, $J$, of a DM halo is typically estimated by using the spin parameter that was originally defined as  \citep[][]{peebles},  
\begin{equation}\label{eq:pe}
\lambda_P = J \frac{E_{\rm DM}^{1/2}}{G M_{\rm DM}^{5/2}} \ ,
\end{equation} 
where $E_{\rm DM}$ is the binding energy associated with the DM halo with mass $M_{\rm DM}$, and $G$ is the gravitational constant. For practical purposes, a more feasible measure of the angular momentum was defined by \citet{bullock} as
\begin{equation}\label{eq:Bu}
\lambda = \frac{j_{\rm sp}}{\sqrt{2}v_{200}R_{200}} \ ,
\end{equation}
where 
\begin{equation}
\mathbf{j}_{\rm sp} = \frac{\mathbf{J}}{M} = \frac{1}{M}\sum_{i=1}^N m_i \mathbf{r}_i \times \mathbf{v}_i \ ,
\end{equation}
$j_{\rm sp}=|\mathbf{j}_{\rm sp}|$, $v_{200} = \sqrt{GM_{200}/R_{200}}$, and $M = \sum_i^N m_i$. The sums are over any set of particles $\{\rm gas, DM, total\}$. The expression in Eq.~(\ref{eq:Bu}) allows for the definition of separate spin parameters for the DM and gas component of a halo. Non-radiative hydrodynamical simulation studies have shown that the spin parameter of the gas component is typically higher than that of the DM component \citep[e.g.,][but see \citealt{vandenbosch02} for a different conclusion]{sharma2005,ZS}. Below we adopt Eq.~(\ref{eq:Bu}) as our definition of the spin parameter. We note that the two alternative definitions for the spin parameter are not entirely equivalent. Analytically, for NFW haloes, the different definitions are related to each other via  $\lambda_P = f(c)^{1/2}\lambda$, where $f(c)$ depends on the concentration $c$ of the halo \citep{mo}.
The spin parameter is well described by a log-normal distribution \citep[e.g.,][]{bullock}
\begin{equation}
P(\lambda) = \frac{1}{\lambda\sqrt{2\pi}\sigma}\exp\left(-\frac{\log^2(\lambda/\lambda_0)}{2\sigma^2}\right) \ ,
\end{equation}
where the fit parameters are:  $\lambda_0=0.035$ and $\sigma=0.5$ \citep[][similar fits were found by \citealt{hetznecker06,ZS,RG}]{bullock}.
In this work, we focus on smaller mass structures $\lsim10^{8}~\msun$, and separate our discussion to the DM haloes with a gas component and Gas-Primary objects. We assume spherical symmetry and the same normalization for each of the halo's component (i.e,. $M_{200}$ and $v_{200}$).

\subsection{DM-Primary/Gas-Secondary Objects}
Recall that DM-Primary/Gas-Secondary objects were found via the FOF algorithm. Using a spherical overdensity algorithm, one can compute an associated virial radius. We then restrict to the spherical sverdensity halo and study its properties. These haloes in general contain both DM and gas particles. Note that the \citet{bullock} definition of the spin parameter (Eq. \ref{eq:Bu}) is appropriate for such haloes given their inherent spherical symmetry assumption. 
 
As depicted in the top panels of Figure \ref{lognormaldist}, the spin parameter distribution of these small-size, high-redshift ($z=10,~20$), DM haloes ($\leq 10^8~\msun$) follows a log-normal distribution, consistent with massive haloes at low redshift. Both the zero and non-zero stream velocity cases give similar results for the DM component in the halo. The fit parameters of the distribution are given in Table \ref{table:Spin}, and they also agree with the low-redshift, massive DM haloes parameters \citep[e.g.,][]{bullock}.

 In the bottom panels of Figure \ref{lognormaldist}, we display the spin parameter of the gas components of these haloes, for $v_\bc=0$ (dashed red lines) and $v_\bc=2\sigma_\bc$ (dotted blue lines). As before, the two representative examples are for $z=10$ and $z=20$. We note first that the gas component distribution of the $v_\bc=0$ case, has a slight evolution in time (see Table \ref{table:Spin}). This evolution is to be expected since  more gas is being accreted onto the DM haloes as the Universe evolves.  Notably, while the stream velocity has a negligible effect on the DM, it has a significant effect on the spin parameter distribution of the gas component. The  gas spin parameter distribution has become flattened leading to a larger median. In comparison to the DM, the gas has spun up by an amount $\lambda_{\rm gas}/\lambda_{\rm DM}$ of  $3.5$ at $z=20$ and of $2.7$ at $z=10$ in the presence of the stream velocity. As time goes by, the DM angular momentum torques the gas angular momentum and re-aligns it. 

\begin{table}
\begin{center}
\begin{tabular}{|c||cc|cc|} 
 \hline
 $\lambda_{\rm DM}$ & \multicolumn{2}{|c|}{$z=20$} & \multicolumn{2}{|c|}{$z=10$} \\
 \hline
 \hline
 & ${\lambda}_0$ & $\sigma$ & ${\lambda}_0$ & $\sigma$ \\
 \hline
 $v_\bc=0$ & 0.04 & 0.61 & 0.04 & 0.68\\
 $v_\bc=2\sigma_{v_\bc}$ & 0.04 & 0.62 & 0.04 & 0.70 \\
 \hline
 \hline
 \hline
 $\lambda_{\rm gas}$ & \multicolumn{2}{|c|}{$z=20$} & \multicolumn{2}{|c|}{$z=10$} \\
 \hline
 \hline
 & ${\lambda}_0$ & $\sigma$ & ${\lambda}_0$ & $\sigma$ \\
 \hline
 $v_\bc=0$ & 0.04 & 0.65 & 0.06 & 0.82 \\
 $v_\bc=2\sigma_{v_\bc}$ & 0.12 & 0.70 & 0.11 & 0.82 \\
 \hline
 \hline
 \hline
 $\lambda_{\rm ellip}$ & \multicolumn{2}{|c|}{$z=20$} & \multicolumn{2}{|c|}{$z=10$} \\
 \hline
 \hline
 & ${\lambda}_0$ & $\sigma$ & ${\lambda}_0$ & $\sigma$ \\
 \hline
 $v_\bc=0$ & 0.07 & 0.68 & 0.07 & 0.74 \\
 $v_\bc=2\sigma_{v_\bc}$ & 0.12 & 0.65 & 0.11 & 0.73 \\
\hline
\end{tabular}
\end{center}
\caption{Lognormal fit parameters for DM component, the gas inside the DM haloes, and the gas ellipsoid structures.}\label{table:Spin}
\end{table}

\begin{figure*}
\includegraphics[width=.45\textwidth]{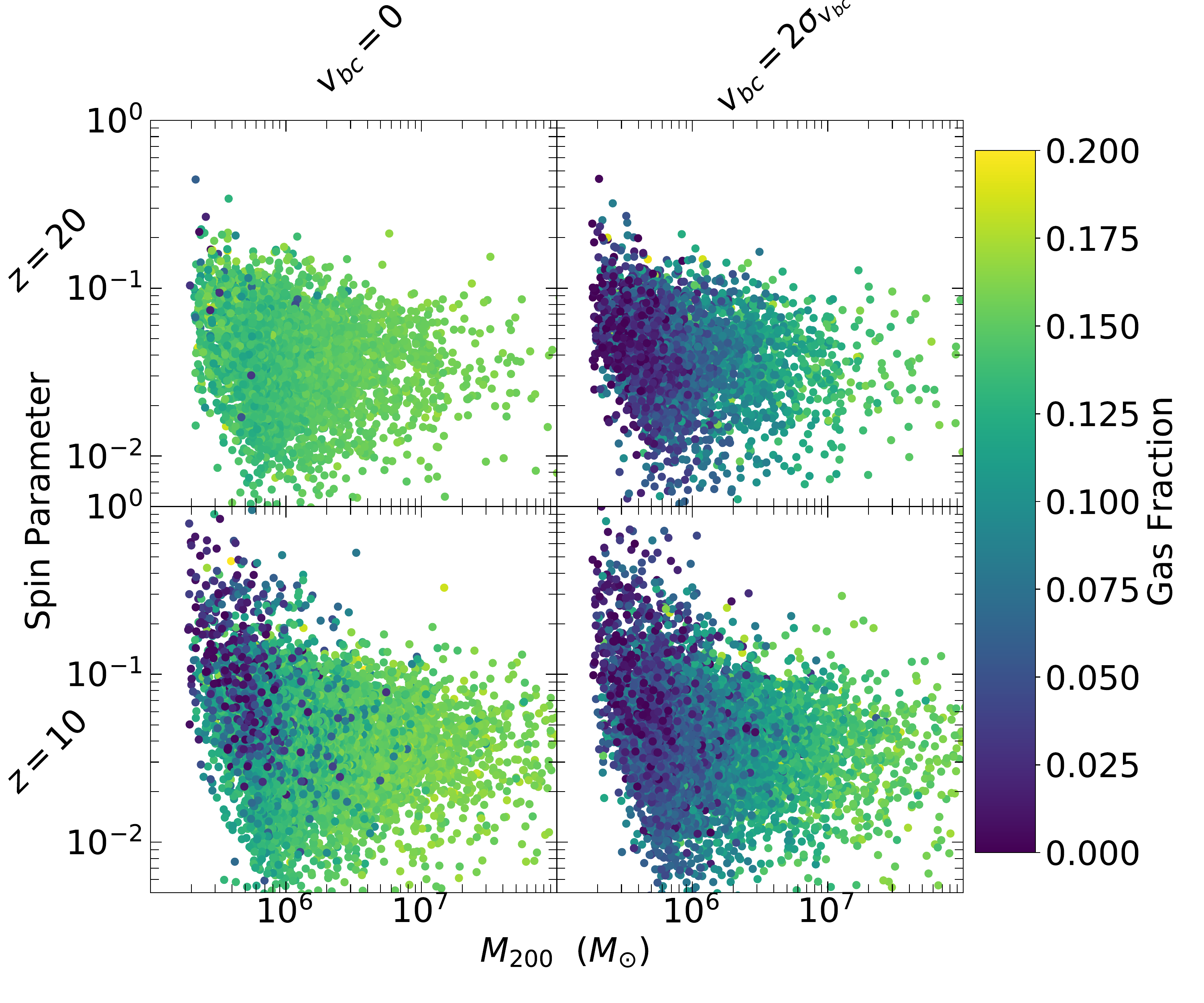}
\includegraphics[width=.45\textwidth]{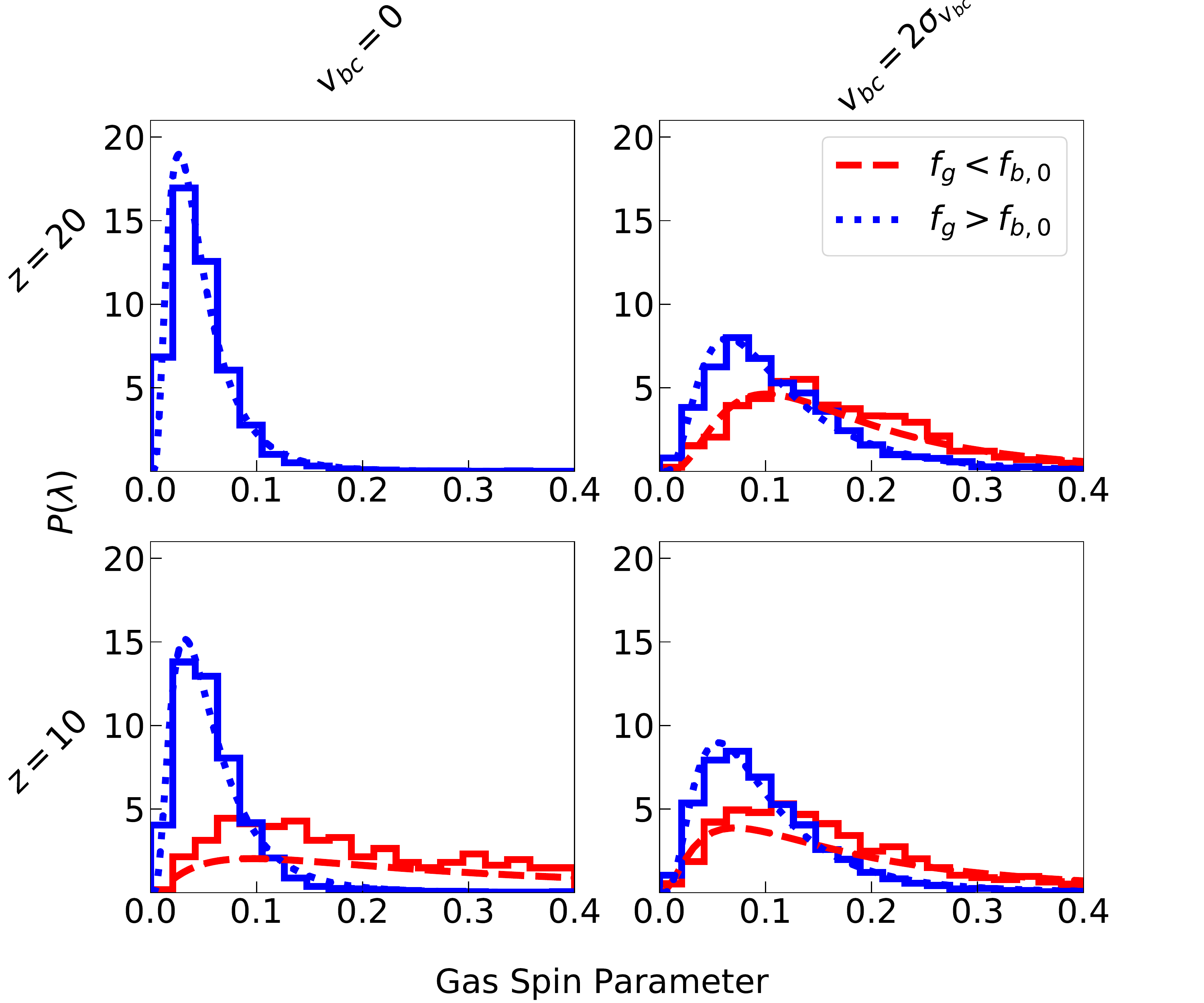}
\caption{The left panels depict distributions of halo spin parameters (including both DM and gas components) as a function of halo mass colored by gas fraction. The right panels show the (lognormal) distributions where we have separated the populations between gas poor ($f_g<f_{\rm b,0}$, dashed red lines) and gas rich ($f_g>f_{\rm b,0}$, dotted blue lines) structures, where $f_{\rm b,0}$ is the gas fraction of the top 5\% massive haloes realized. The stream velocity effect increases the median spin parameter more for gas poor haloes than gas rich haloes.}
\label{totspinparam}
\end{figure*}

Smaller mass haloes are more sensitive to stream velocity effects. This is depicted in the left panels of Figure \ref{totspinparam}, where we present the combined spin parameter of both gas and DM as a function of the total halo mass. The first row is for $z=20$ while the second is for $z=10$. The colors represent the gas fraction in each halo. As expected, the stream velocity decreases the gas fraction of a halo \citep[e.g.,][]{popa}. In the right panels of Figure \ref{totspinparam}, we plot the gas component of the spin parameter but split the objects into those that were above and below $f_{\rm b,0}$ to distinguish between gas-rich and gas-poor haloes. This shows that the gas spin parameter distribution of gas-poor haloes is more extended in the stream velocity case as compared to the zero stream velocity simulation. We see that the objects below the threshold in $f_{\rm b,0}$ exhibit slightly higher median gas spin parameter. The gas content in these objects is smaller due to the presence of the stream velocity, which corresponds to higher spin parameter for the gas-poor haloes. Gas-rich haloes represent those objects that are less sensitive to $v_\bc$, i.e., typically  DM halo mass larger than a few$~\times 10^6$~M$_\odot$ \citep[e.g.,][]{naoznarayan14}, and thus results in the nominal,  $v_\bc=0$ spin parameter peak value, (see blue lines in the left panels and green dots in the right panels in Figure \ref{totspinparam}). On the other hand, smaller mass objects are more sensitive to $v_\bc$ and thus, are gas deficient and have higher spin parameter peak value, compared to the gas-rich ones  (see red lines in the left panels and dark blue dots in the right panels in Figure \ref{totspinparam}). In the nonzero stream velocity case, the gas-poor haloes accumulate gas preferentially from the stream flow which causes the excess of angular momentum compared to the $v_\bc=0$ case.

\begin{figure}
\includegraphics[width=.4\textwidth]{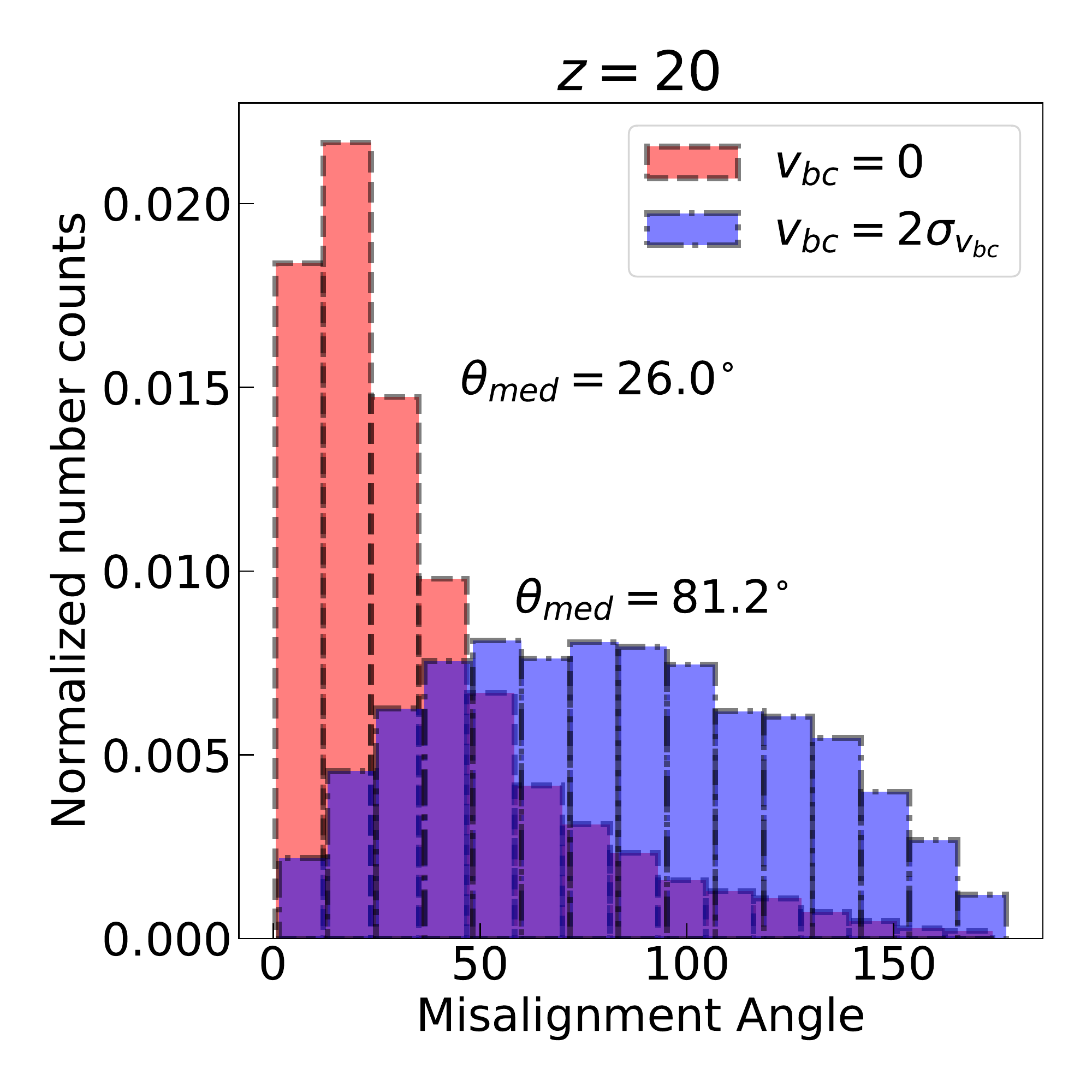}
\includegraphics[width=.4\textwidth]{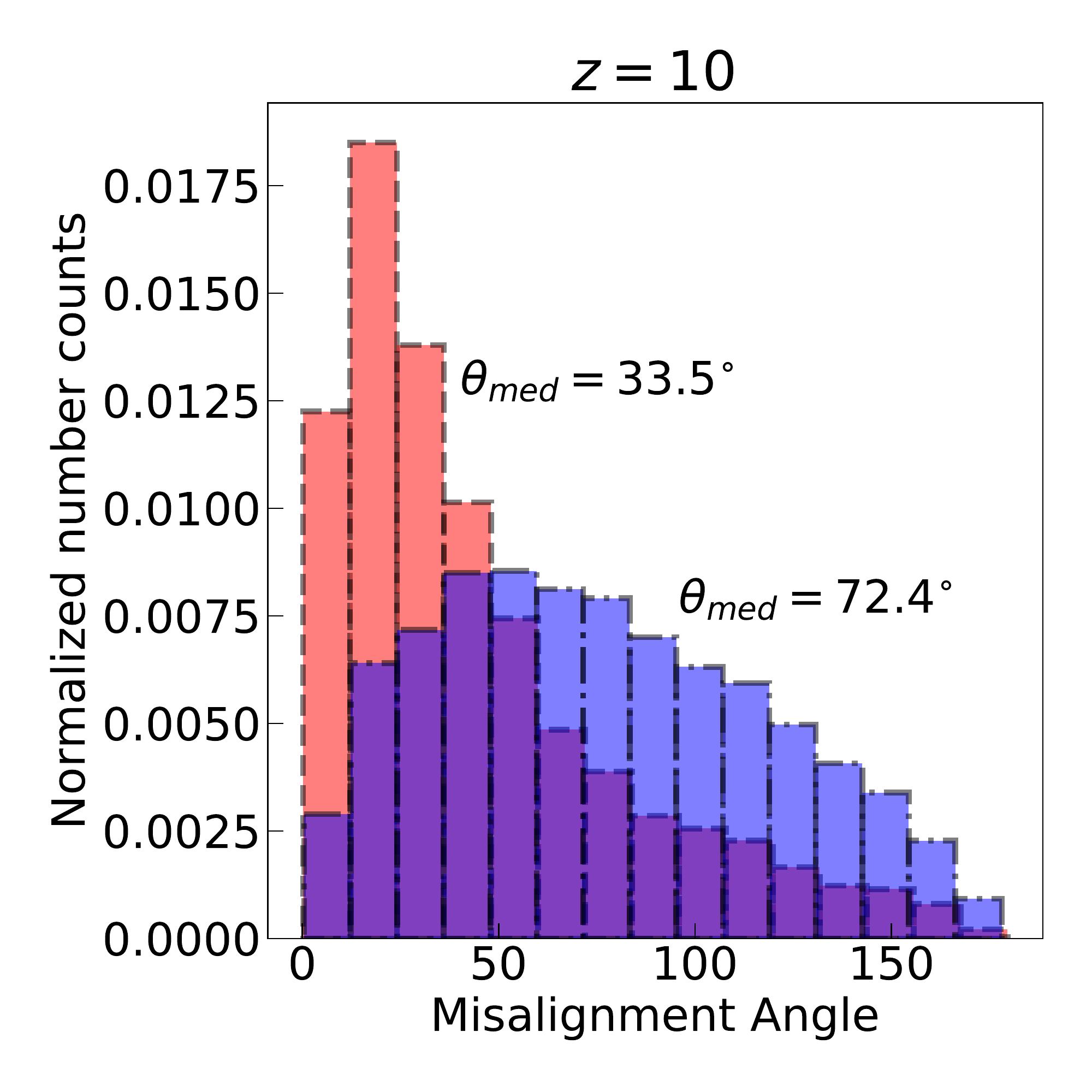}
\caption{Histogram of misalignment angle between the DM and baryonic component (see Eq.~\ref{eq:mis}) . We consider the zero $v_{\rm bc}$ case in dashed red and the $v_{\rm bc}=2\sigma_{v_{\rm bc}}$ in dot-dashed blue. The $v_{\rm bc}=0$ result is in agreement with previous studies \citep{vandenbosch02,ZS}. Note the nearly isotropic distribution at early times with stream velocity (i.e., $z=20$). }\label{misalignmentdist}
\end{figure}

We calculate the misalignment angle $\theta$ between the DM and gas particle angular momentum:
\begin{equation}\label{eq:mis}
\cos(\theta) = \frac{\bm{J}_{\rm DM}\cdot \bm{J}_{\rm gas}}{|J_{\rm DM}||J_{\rm gas}|} \ .
\end{equation}
Previous simulation studies have shown that there is a moderate misalignment of $\sim 30^{\circ}$ between the DM and baryons \citep[e.g.,][]{vandenbosch02,ZS}.
As depicted in Figure \ref{misalignmentdist}, we find a similar median misalignment angle in the zero stream velocity case as in previous studies. However, the stream velocity effect results in a nearly isotropic distribution of the misalignment angle and increases the median to about $81^{\circ}~(72^{\circ})$ at $z=20$ ($z=10$). The DM did not have enough time to torque the gas component into alignment as the gas accretes at later times onto these haloes.

\subsection{Gas-Primary and Supersonically-Induced Gas Objects}

We define a Gas-Primary object by running the FOF algorithm only on the baryonic component and tightly fitting an ellipsoid around them (see Section \ref{sec:struc}). We also note that some of these structures are, in fact, the gas component in a DM halo. We refer to those that lie beyond the virial radius of their parent dark matter halo and also have a gas fraction larger than $40\%$ as SIGOs. 

\begin{figure}
\centering
\includegraphics[width=.4\textwidth]{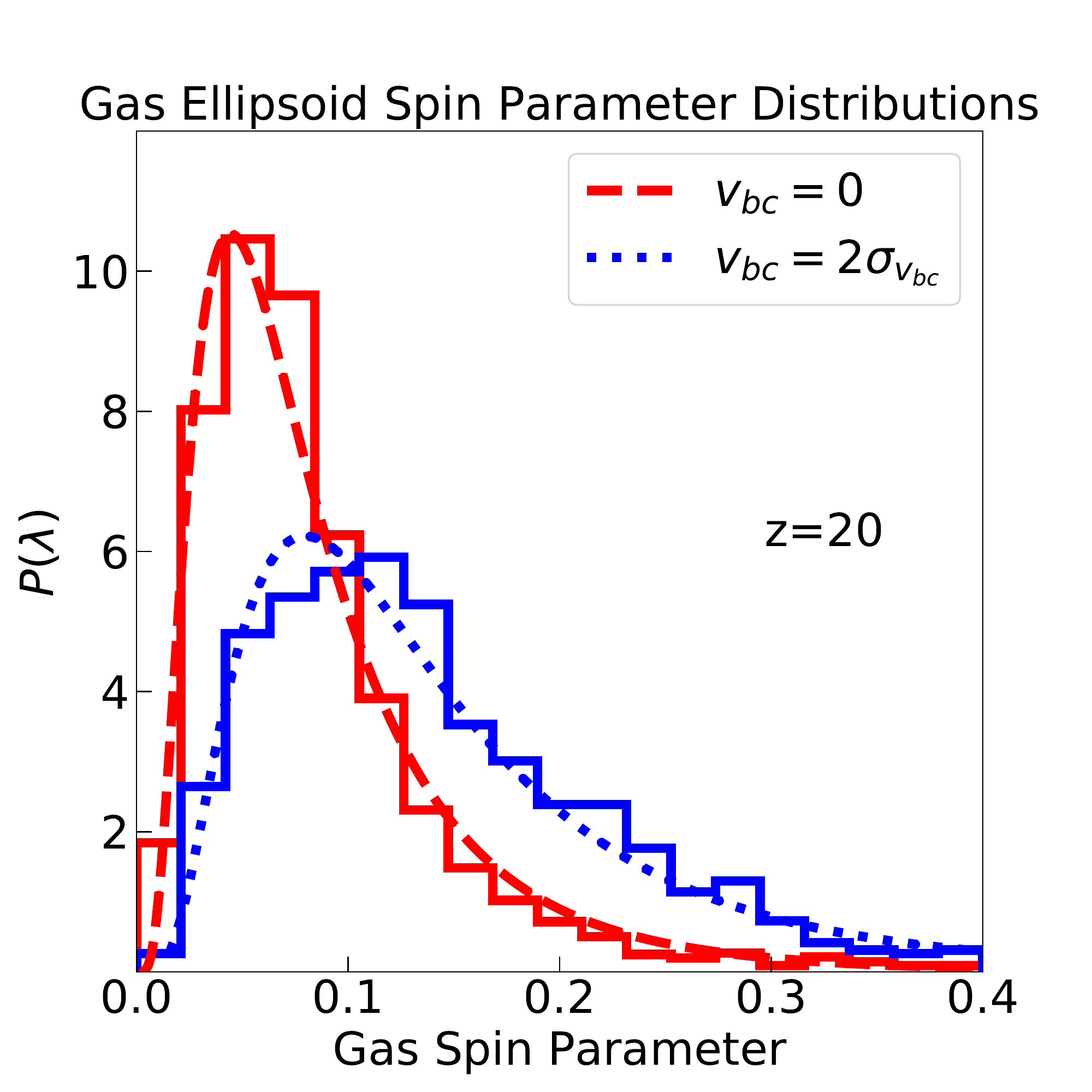}
\includegraphics[width=.4\textwidth]{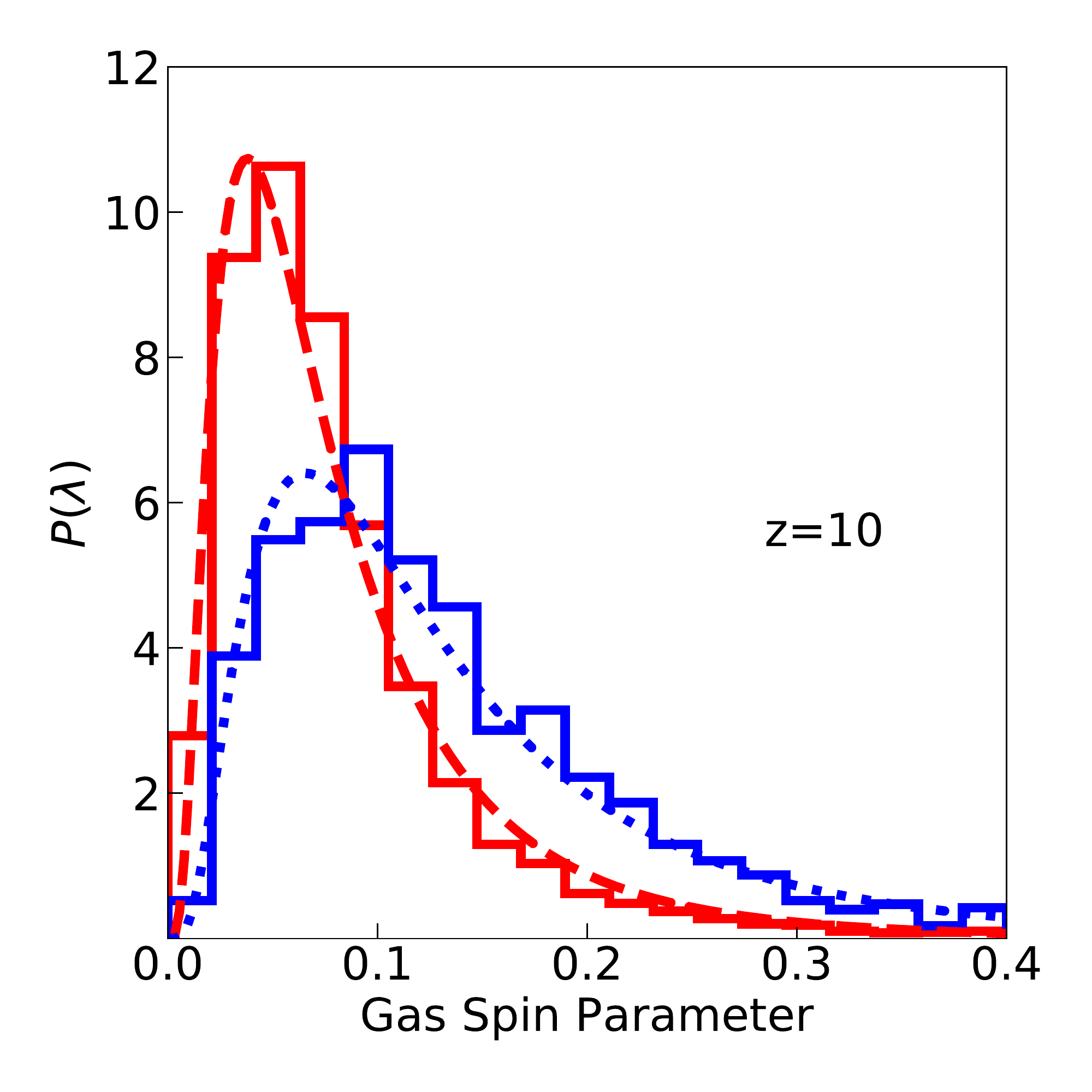}
\caption{Lognormal distribution of generalized spin parameters for Gas-Primary objects at different redshifts. We consider the zero $v_{\rm bc}$ case in dashed red and the $v_{\rm bc}=2\sigma_{v_{\rm bc}}$ in dotted blue.}
\label{GasEllipsoidlognormaldist}
\end{figure}

We extend the spin parameter definition to the Gas-Primary structures. Since these structures are ellipsoidal in nature \citep[e.g.,][and see Figure \ref{gasproj}]{popa}, the spherical symmetry associated with the spin parameter normalization does not hold. Furthermore, as shown in Figure \ref{gasproj} the gas inside the Gas-Primary structures does not have an NFW distribution. In particular, Gas-Primary structures, tend to have a flatter inner distribution, unlike the DM haloes. Therefore, in order to compare between the DM-Primary/Gas-Secondary and Gas-Primary (and SIGOs), we need to choose a self-consistent normalization. We choose a normalization in accordance with the physics picture described in \citet{naoznarayan14}, in which Gas-Primary structures originated from the corresponding overdensity mode of their DM counterparts. These are simply defined by $\sim 6\times M_{\rm gas}$, where $M_{\rm gas}$ is the mass of the gas structure. Furthermore, as it can be seen in Figure \ref{gasproj}, the complex dynamical evolution of these systems, resulted in having the gas-rich clumps concentrated around a larger haloes. In other words they have not stayed with the parent DM halo. Hence, we define a spin parameter for the ellipsoidal Gas-Primary objects as 
\begin{equation}
\lambda_{\rm GP} = \frac{J_{\rm gas}}{6\sqrt{2} M_{\rm gas} v_{{\rm GP}} R_{{\rm max}}} \ ,
\end{equation}
where $R_{\rm max}$ is half of the longest axis of the fitted ellipsoid, and $v_{{\rm GP}}$ is the circular velocity at this distance. 
This choice of similar normalization for the Gas-Primary objects, allows us to have one-to-one comparison with the definition of the spin parameter (Eq. \ref{eq:Bu}). As shown in Figure \ref{GasEllipsoidlognormaldist}, the extended spin parameter definition also follows a lognormal distribution and the stream velocity effect shifts the median higher. This is expected because a majority of the Gas-Primary objects are just the gas components of a corresponding DM-Primary/Gas-Secondary object.

\begin{figure}
\includegraphics[width=0.5\textwidth]{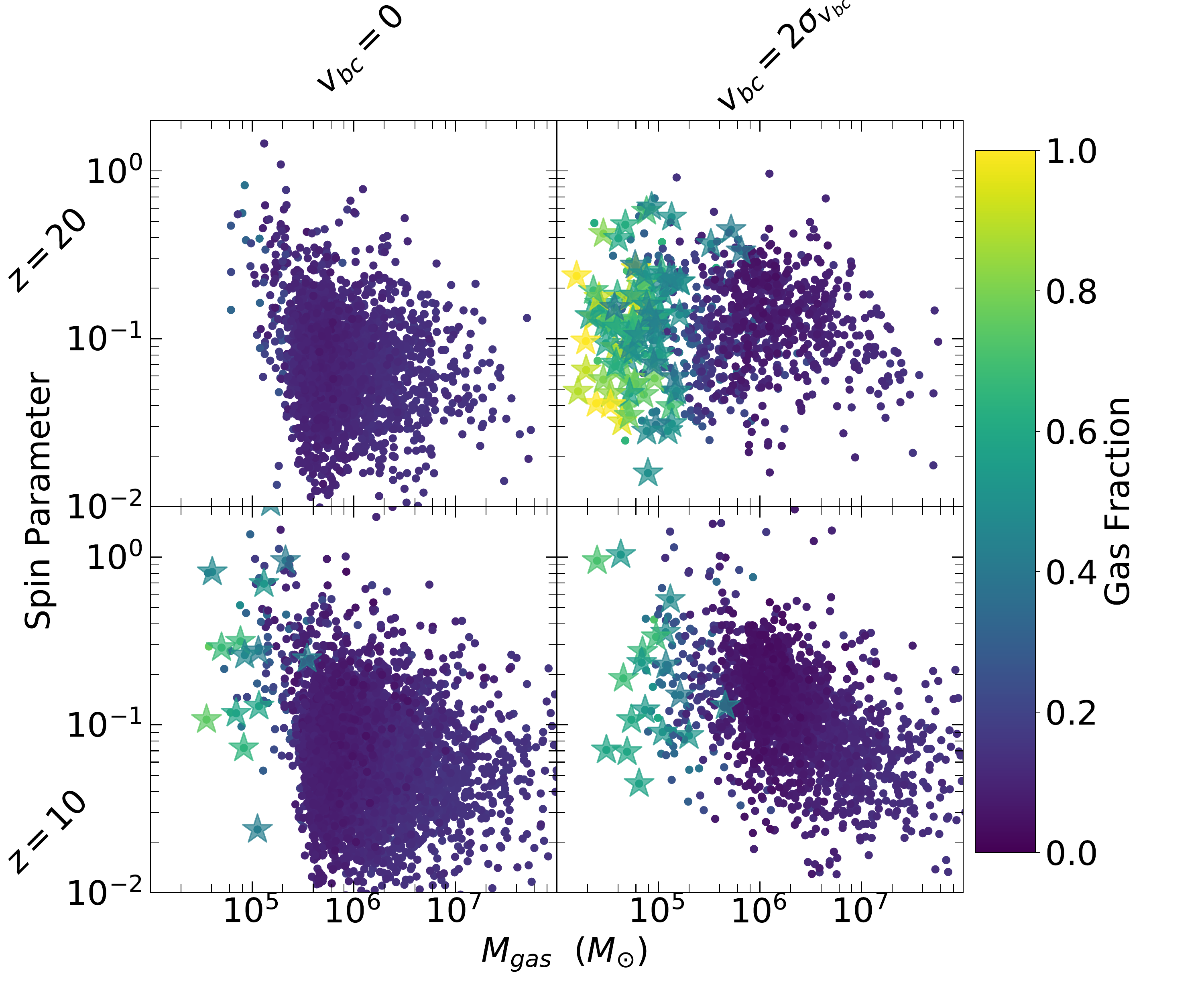}
\caption{Spin parameter distributions for Gas-Primary objects with and without stream velocity at different redshifts. We mark the Supersonically-Induced Gas Objects (SIGOs) as stars.  Note that SIGOs occupy a large range of spin parameters, and as predicted in \citet{naoznarayan14}, are typically below  $10^6$~M$_\odot$. The color code describes the gas fraction in these structures.  }
\label{GasEllipsoidspinparam} \vspace{0.2cm}
\end{figure}

Similar to our presentation of the spin-mass relationship of the DM-Primary/Gas-Secondary objects (see Figure \ref{totspinparam}), we analyze the spin-mass behavior of the Gas-Primary objects in Figure  \ref{GasEllipsoidspinparam}. We mark the SIGO subset of objects in stars. As shown in the Figure, SIGOs have typically smaller masses \citep[as expected from the analytical analysis,][]{naoznarayan14}, and systematically larger spin parameter.  

\begin{figure}
\centering
\includegraphics[width=0.5\textwidth]{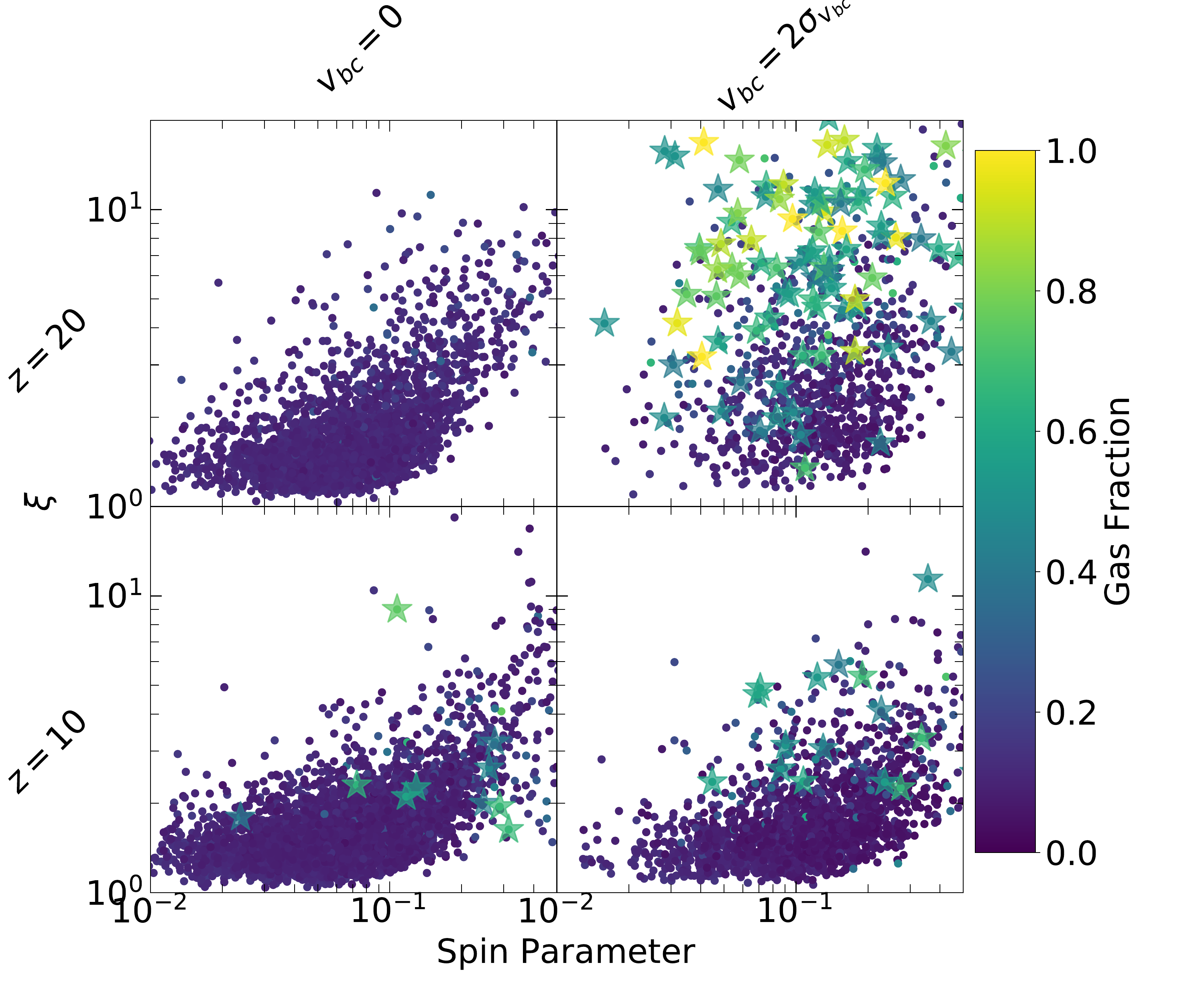}
\caption{The prolateness of ellipsoids (see Eq.~\ref{eq:pro}) as a function of spin parameter. $R_{\rm max}$ and $R_{\rm min}$ refer to the largest and smallest axis of the ellipsoid in question. Stars refer Supersonically-Induced Gas Objects (SIGOs). The colour code indicates the gas fraction in each object. Note that $\zeta\to 1$ means a spherical system. Rotationally supported systems have large prolateness and large spin parameter.}
\label{spinparamvsRmaxRmin}
\end{figure}

Another question one can ask is how the ellipsoidal morphology is affected by the spin parameter. Under the stream velocity effect, the ellipsoidal Gas-Primary structures, (see Figure \ref{spinparamvsRmaxRmin}), become more prolate. There is also a small positive correlation between the prolateness and the spin parameter that decreases with redshift. We suspect that this correlation exists because the baryonic components are collapsing on to disks, which gives higher spin. However, with the stream velocity effect included, this correlation disappears at high redshift, where the stream velocity has a greater effect. The weak correlation is restored at low redshift, but in this case the objects are prolate ellipsoids in shape. The SIGOs, however, do not seem to show any correlation between spin parameter and prolateness. 

\section{Summary and Conclusions} \label{sec:conc}

The supersonic relative velocity between the DM and baryons in the early Universe has been shown to shift the positions of baryonic and DM over-densities \citep[e.g.,][]{naoz12}, and in some cases the baryonic over-densities may collapse outside the virial radius of their parent DM halo, forming the so-called \textit{Supersonically-Induced Gas Objects} \citep[e.g.,][]{naoznarayan14,popa}. We identified two significant structures (see Section \ref{sec:numerics}). The first structures are the classical DM haloes, with a gas component. These were found in the simulation by using a DM-Primary/Gas-Secondary linking in the  FOF groups. The other structures were identified in the simulation using only gas particles in the FOF groups, namely the Gas-Primary objects. The stream velocity effect further produces SIGOs (which are a subset of Gas-Primary objects). This new class of objects is defined by lying outside the virial radius of DM haloes and also have a gas fraction of over $40\%$. At $z=20$, $\sim 11$\% of the gas-rich objects ($f_g > f_{\rm b,0}$) are in fact SIGOs.  

We have investigated the angular momentum of all preceding objects through the spin parameter. To our knowledge, this is the first study of spin parameters of simulated haloes at the low mass range ($10^{5-8}~\msun)$ at high-redshift.

We began by investigating the DM haloes and their baryonic counterparts. We found that the spin parameter of the DM haloes is described by a lognormal distribution, just as the low-redshift, high-mass counterparts. Moreover, the fit parameters are consistent with those obtained for high-mass, low redshift haloes (see Table \ref{table:Spin}, and see for example, \citealt{bullock}). The relative velocities between baryons and DM produce non-trivial consequences to the spins parameters of these objects. In particular, the stream velocity effect increases the median spin parameter of the gas components of DM haloes, as seen in Figure \ref{lognormaldist}.

We have demonstrated in Figure \ref{misalignmentdist} that the angular momenta of the gas and DM components are highly misaligned, due to the stream velocity effect. The  angle between the angular momentum vectors of the DM and gas component of the halo has a nearly isotropic distribution.  The standard picture of galaxy formation posits that both baryons and DM inherit the same initial angular momentum from tidal torques. However, baryonic effects in the case of zero stream velocity still produce about $30^\circ$ misalignment \citep[e.g.,][]{vandenbosch02,velliscig15,liao17,Jiang+18}.
Notably, the stream velocity effects yields a larger misalignment that can occur in high-redshift galaxies. In particular, at high redshift the misalignment angle roughly follows an isotropic distribution.  

We then generalized the spin parameter to accommodate the ellipsoidal Gas-Primary objects which also include the new class of Supersonically-Induced Gas Objects (SIGOs) that arise from the stream velocity effect. The spin parameter traditionally normalizes the angular momentum to with the virial parameters of the DM halo. Since SIGOs do not live inside a DM halo\footnote{Note that the gas ellipsoids with gas fraction lower then $40\%$ are dominated by DM, but not necessarily in the centre of a DM halo \citep{naoznarayan14}.}, we normalize spin parameters for Gas-Primary objects to a fiducial host DM halo of mass $6M_{\rm gas}$. 
The majority of the Gas-Primary objects (90\% at $z=20$) are gas components of DM haloes, as such they are expected to follow a log normal distribution as expected \citep[e.g.,][]{bullock,ZS} The lognormal distribution in the classical objects are shown in Figure \ref{lognormaldist}, and the log-normal distribution of the Gas-Primary ellipsoids is demonstrated in Figure \ref{GasEllipsoidlognormaldist}. Thus, although these are classified differently, they represent the same component. Focusing on the SIGOs, we found that their morphology, in particular, their prolateness is uncorrelated with the spin parameter, in contrast to typical disk galaxies which would exhibit a positive correlation (as depicted in Figure \ref{spinparamvsRmaxRmin}).

We showed that in the presence of stream velocity, even in the non-radiative case, the spin parameter the gas component of DM haloes is increased, and is misaligned with respect to the DM's angular momentum. Moreover, we showed that the SIGOs have higher spin parameters in the presence of the $v_{\rm bc}$ compared to Gas-Primary in the case of $v_{\rm bc}=0$. The spin parameter already serves as a valuable measurable quantity for higher mass galaxies ($\lsim 10^{10}$~M$_\odot$), and lower redshift ($z\sim 1-3$) formation histories \citep[e.g.,][]{burkert16}. 

These SIGOs are likely to form stars from their gas concentrations and are typically small \citep [about a few kpc, which expected to be even smaller when cooling is introduced e.g.,][]{hirano17b}. Therefore they might explain the low luminosity, small-sized objected recently reported by \citet[][]{Bouwens+17}. Furthermore, 
JWST may be able to detect SIGOs and low luminosity DM haloes at their formation sites, $z\sim 20$. The large spin parameter, if inferred from observational data, might serve as an indication for a high stream velocity patch. Recently, HST has observed a potentially promising candidate, the so-called ``Little Blue Dots'' \citep{elmegreen17}. These objects are thought to be gas-dominated systems in the early Universe, with high specific star formation rates. Based on the gas fraction and the scale of the structures, we speculate SIGOs are a possible formation channel for these observed objects.

\section*{Acknowledgements}
We thank Volker Springel for access to {\sc arepo}. We thank Joel Primack for insightful questions about the spin parameter that motivated this project. We also thank Ramesh Narayan, Re'em Sari, Rachel Somerville, James Bullock, and Michael Boylan-Kolchin for enlightening discussions. We also thank Cristina Popa for her help in the early stages of this project. 

\bibliographystyle{mnras}
\bibliography{main}

\bsp
\label{lastpage}
\end{document}